\title{On the Hamming distance between base-$n$ representations of whole numbers}
\author{
        Anunay Kulshrestha \\
                Delhi Public School, Dwarka\\
      New Delhi 110 075\\
}
\date{January 21, 2012}
\documentclass[12pt]{article}
\usepackage{amsfonts}
\usepackage{amssymb}

\begin{document}
\maketitle

\begin{abstract}
We first introduce the Hamming distance between two strings. Then, we apply this concept to the representations of whole numbers in base $n$ for all positive integers $n > 2$. We claim that a simple formula exists for the sum of all Hamming distances between pairs of consecutive integers from 1 to $m$, which we will derive. We also state and prove other interesting results concerning the aforementioned topic.
\end{abstract}

\section{Introduction}
\paragraph{}
The remainder of this paper is organized as follows : 
Section $2$ provides an introduction to the Hamming distance and its applications.
Some new results are stated with proof in Section $3$.
Section $4$ gives a Python implementation of the algorithm developed.. It also carries programs in Python for the same. 
Finally, Section $5$ concludes the paper.

\section{Hamming Distance}
The Hamming distance is named after Richard Hamming. Defined between two strings of equal length, the Hamming distance gives the number of positions at which corresponding symbols differ. Alternatively, it measures the minimum number of substitutions required to change one string into the other, or the number of errors in transmission, not including insertions and deletions, required to transform one string into the other. In this paper, we will use the notation $H(S_1, S_2)$ to denote the Hamming distance between two strings of equal length - $S_1$ and $S_2$.
\newline
\newline
For example, the \verb|H(math, mats) = 1| while \verb|H(math, math) = 0|. Hamming distance is not defined between two strings of unequal length.
\newline
\newline
The following Python function calculates the Hamming distance, taking two strings of equal length as arguments :

\begin{verbatim}
def hammingdistance(s1, s2):
    assert len(s1) == len(s2)
    return sum(ch1 != ch2 for ch1, ch2 in zip(s1, s2))
\end{verbatim}

\section{Results}\label{results}
Let $n \in \mathbb{N}$  and $n \ge 2$. We consider the representations of whole numbers in base $n$. We append zeros to the left of the numerals in case the strings are of unequal length. This happens between $m-1$ and $m$ if and only if $m$ is a power of $n$.
\newline
\newline
{\underline {Lemma}} : $H(m, m-1) = l+1$ where $l$ is the exponent of the highest power of $n$ that divides $m$.
\newline
\newline
{\underline {Proof}} : We prove this by considering $3$ cases.
\newline
\newline
Case $1$ : $m = n^l$ for some $l \in \mathbb{N}$
\newline
\newline
Here, $m = {(100...000)}_n$ ($1$ followed by $l$ zeroes).
Let A be the digit in base $n$ that represents $n - 1$. So, $m - 1 =$ AAA...AAA ($l$ A digits).

Clearly, $H(m, m-1) = l + 1$ and we are done.
\newline
\newline
Case $2$ : $m = pn^q$ for some $q \in \mathbb{N}$ where $q \ge 1$ and $(p, n) = 1$. Let,

\begin{equation}
m = {n^k}{a_k} + {n^{k-1}}{a_{k-1}} + ... + {n^2}{a_2} + {n^1}{a_1} + {n^0}{a_0}
\end{equation}

Trivially, all $a_j$ for $j < q$ will be zeros. 

And, 

\begin{equation}
m - 1 = {n^k}{b_k} + {n^{k-1}}{b_{k-1}} + ... + {n^2}{b_2} + {n^1}{b_1} + {n^0}{b_0}
\end{equation}

Here, $a_0 = 0$ as $n$ divides $m$.

Now, $m - 1$ will have all $b_j = (n-1)$ for $j < q$ and $b_q = a_q - 1$. For $i > q$, on the other hand, $a_i = b_i$.
\newline
\newline
Thus, $H(m, m-1) = q + 1$ and we are done.
\newline
\newline
Case $3$ : $n$ does not divide $m$. Let
\begin{equation}
m = {n^k}{a_k} + {n^{k-1}}{a_{k-1}} + ... + {n^2}{a_2} + {n^1}{a_1} + {n^0}{a_0}
\end{equation}
\begin{equation}
m = \sum\limits_{i=0}^k {n^i}{a_i}
\end{equation}

Similarly,

\begin{equation}
m - 1 = {n^k}{b_k} + {n^{k-1}}{b_{k-1}} + ... + {n^2}{b_2} + {n^1}{b_1} + {n^0}{b_0}
\end{equation}
\begin{equation}
m - 1 = \sum\limits_{i=0}^k {n^i}{b_i}
\end{equation}
Since $n$ does not divide $m$, $a_0 \geq 1$. Therefore $b_0$, the remainder of $m - 1$ upon division of $n$, is simply $a_0 - 1$, and $b_i = a_i$ for all $i \geq 1$. Therefore $H(m, m-1) = 1$ and we are done. \hfill${\blacksquare}$
\newline
\newline

Now that we have found a way to calculate the Hamming distance between two consecutive numbers in base $n$, we will derive a formula for the sum of all such Hamming Distances up to a given $m$. Let
\begin{equation}
S(m) = H(0, 1) + H(1, 2) + H(2, 3) ... + H(k-2, k - 1) + H(k - 1, k)
\end{equation}
\begin{equation}
S(m) = \sum\limits_{i=0}^{k-1} H(i, i+1)
\end{equation}

From the lemma, we know that $H(m, m-1) = l+1$ where $l$ is the exponent of the highest power of $n$ that divides $m$.

Thus,
\begin{equation}
S = \sum\limits_{i=1}^{m} (P_i + 1)
\end{equation}

Where $P_i$ represents the highest power of $n$ contained in $i$. Trivially, the least value that $P_i + 1$ will ever attain for any $i$ is $1$. Using Iverson Brackets,
\newline
\begin{equation}
S(m) = \sum_{i=1}^m \sum_{j=0}^\infty [n^j | i]
\end{equation}

which we can then rewrite as, by exchanging the order of summation,

\begin{equation}
S(m) = \sum_{j=0}^\infty \sum_{i=1}^m [n^j | i]
\end{equation}

But there are $\lceil m/n^j \rceil$ integers from 1 to $n$ that divide $n^j$, so

\begin{equation}
S(m) = \sum_{j=0}^\infty \lceil m/n^j \rceil
\end{equation}

Finally, we only need to take the sum up to $\lceil \log_n m \rceil$, since $m < n^j$ for $j > \log_n m$.
\newline

And we are done. \hfill $\blacksquare$

\section{Implementation}
The following Python code finds the sum of Hamming distances in base $n$ upto a given $m$ :

\begin{verbatim}
i = int(raw_input('Enter a number : '))
n = int(raw_input('Enter a base : '))

i = int(i, n)

s = 0
x = 0

def hamming_distance(s1, s2):
    assert len(s1) == len(s2)
    return sum(ch1 != ch2 for ch1, ch2 in zip(s1, s2))



while (x < i):
    s1 = base10toN(x, n)
    s2 = base10toN(x+1, n)
    if (len(s1) != len(s2)):
         s1 = '0' + s1
    s = s + hamming_distance(s1, s2)
    x = x + 1

print s
\end{verbatim}

The following Python code prints the sum of all Hamming distances from 0 to m in base n using the results proved in this paper : 

\begin{verbatim}
m = int(raw_input('Enter a number : '))
n = int(raw_input('Enter a base : '))

j = i
h = 0
while (j > 1):
    j = j / n
    h = h + 1

s = 0
while (h >= 0):
    s = s + ( i / (n ** h) )
    h = h - 1
	
print s
\end{verbatim}

\section{Conclusions}
We conclude that the sum of the Hamming distances of consecutive integers is given by a simple formula and can be computed easily and efficiently.

\section{References}
\begin{enumerate}
\item Hamming, R. W. Error detecting and error correcting codes. Bell System Tech. J. 29, (1950). 147 - 160.
\item Public Domain Material, Federal Standard 1037C
\end{enumerate}

\bibliographystyle{abbrv}
\bibliography{main}

\end{document}